\documentclass[preprint2]{aastex6}

\newcommand{\Ha}{H$\alpha$}

\newcommand{\Hb}{H$\beta$}

\newcommand{\heI}{He{\sc i}}

\newcommand{\ozone}{He$^{+}$+H$^{+}$}

\newcommand{\nzone}{He$\rm ^{o}$+H$^{+}$}

\newcommand{\tc}{{$\theta^1$~Ori~C}}
\newcommand{\tA}{{$\theta^2$~Ori~A}}

\newcommand{\hii}{H{\sc ii}}
\newcommand{\oiii}{[O~III]}

\newcommand{\neonii}{[Ne~II]}
\newcommand{\neoniii}{[Ne~III]}
\newcommand{\nii}{[N~II]}
\newcommand{\sii}{[S~II]}

\newcommand{\Ne}{n$\rm_{e}$}

\newcommand{\Tstar}{T$\rm_{star}$}
\newcommand{\boldU}{{\bf U}}
\newcommand{\Usky}{{\bf U}$\rm_{sky}$}
\newcommand{\NH}{n$\rm_{H}$}
\newcommand{\rsky}{$\rm r_{sky}$}
 \newcommand{\rtrue}{{\bf r}}
\newcommand{\Qlyc}{$\rm Q_{LyC}$}
\newcommand{\rholyc}{$\rm \rho_{LyC}$}

\begin{document}

\title{Which stars are ionizing the Orion Nebula ?
\footnote{Based on observations from the European Southern Observatory at Paranal, and the Hubble Space Telescope}}

\author{C. R. O'Dell\affil{3}}
\affil{Department of Physics and Astronomy,
Vanderbilt University,
Nashville, TN 37235-1807, USA}

\author{W. Kollatschny\affil{2}}
\affil{Institut f\"ur Astrophysik, Universit\"at G\"ottingen, Friedrich-Hund-Platz 1, 37077 G\"ottingen, Germany}

\and

\author{G. J. Ferland\affil{1}}
\affil{Department of Physics and Astronomy, University of Kentucky, Lexington, KY 40506}

\begin{abstract}
The common assumption that \tc\ is the dominant ionizing source for the Orion Nebula is critically examined.
This assumption underlies much of the existing
analysis of the nebula. In this paper we establish through comparison of the 
relative strengths of emission lines with expectations from Cloudy models and through the direction of the bright
edges of proplyds that \tA, which lies 
beyond the Bright Bar, also plays an important role. \tc\ does dominate ionization in the inner part of the Orion Nebula, but outside of the Bright Bar as far as the southeast boundary of the Extended Orion Nebula, \tA\ is the dominant source. In addition to identifying the ionizing star in sample regions, we were
able to locate those portions of the nebula in 3-D.
 This analysis illustrates the power of MUSE spectral imaging observations
 in identifying sources of ionization in extended regions.

\end{abstract}



\section{Introduction} \label{sec:intro}

The central region of the Orion Nebula (NGC1976) is arguably the best studied \hii\ region \citep{ode08b}. 
As the exemplar of its class of gaseous nebulae it has also been subject to many attempts 
at modeling the physics that occurs and the 3-D structure. We understand that 
it is basically an irregular concave thin layer of ionized gas lying nearer the observer than the Main Ionization Front (MIF) that marks the boundary with the host background molecular cloud. There is also a nearby foreground thin layer (the Veil) of partially ionized gas. 

The usual assumption is that ionization in this region is dominated by the brightest star in the compact Trapezium group, the complex hot star \tc, whose spectral type is usually assigned as about O6. The next hottest and most luminous star is \tA , which lies 135\arcsec\ southeast of \tc. As one
moves further from the Trapezium the degree of ionization decreases, consistent with the ionizing star or stars 
being in the central nebula \citep{ode10}. It is important in interpreting the nebula's spectra to understand if a secondary
star or stars play a role in ionization of the nebula. 

The inner ionized nebula is actually very complex in structure, there being irregular features in the concave surface, the most famous of which is the Bright Bar, and there is a neutral cloud of material containing very young stars known as the Orion-South cloud (hereafter Orion-S) lying 55\arcsec~to the southwest of \tc. Interpreting the spectra and emission-line 
images of the nebula demands knowing the source of ionization for each region. That is the goal of this study.

Two approaches are adopted. In Section~\ref{sec:ionization} we utilize the predictions of key tracers of different
degrees of ionization. In Section~\ref{sec:proplyds} we utilize resolved ionization fronts formed around gas surrounding many of Orion's proplyds. 

Numerous assumptions have been made in this study.
There are many similar values for the distance to the Orion Nebula, but in this study we have adopted 388$\pm$8  pc from
the recent radio results of  \citet{mk16}. For \tc, we have adopted a
temperature (\Tstar) of 38950 K and total luminosity in photons capable of ionizing hydrogen (\Qlyc) of 7.35$\times$10$^{48}$ photons s$\rm ^{-1}$ \citep{Badnell}. For \tA\ at O9.5~V \citep{wh77}, we use values from Table 2.3 of \citet{agn3} of \Tstar~=~34600 K and \Qlyc~=~3.63$\times$10$^{48}$ photons s$\rm ^{-1}$. We express distances in the plane of the sky in parsecs, using the adopted distance.

\begin{figure*}
\figurenum{1}
\plotone{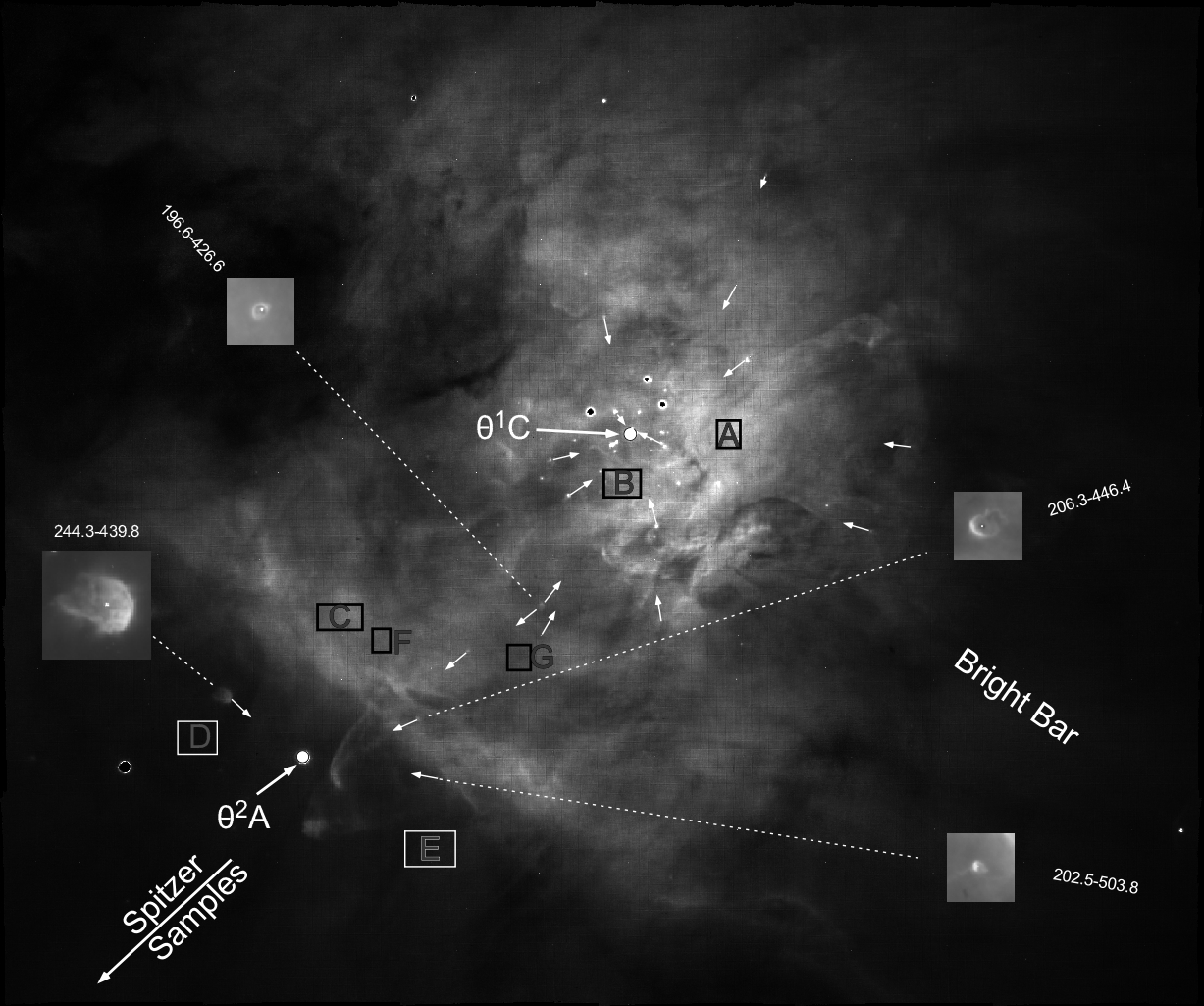}
\caption{
This \Ha\ 0.66$\times$0.55 pc~(353\arcsec $\times$ 295\arcsec) image of the Orion Nebula is from the 0.2\arcsec / pixel MUSE data.
The samples employed in generating the data in Table~\ref{tab:table1} are shown as lettered rectangles the size of the sample. 
The arrows indicate the direction(s) of the orientation of the bright edges of some of the many proplyds as discussed in Section \ref{sec:proplyds}. Those proplyds ionized by \tA\ are shown in enlargments of 2\arcsec $\times$2\arcsec , except for 244.3-439.8 which is a sample of  4\arcsec $\times$4\arcsec. 
The location of the first of the Spitzer Space Telescope observations discussed in Section~\ref{sec:rubin}
are shown south of \tA\ and further defined in Figure \ref{fig:rubin}}
\label{fig:mosaic}
\end{figure*}

\section{Using line ratios to identify the ionizing star}
\label{sec:ionization}

The ionized layer on the observer's side of the MIF is stratified into layers according to the varying ionization. These layers are 
determined by the energy of the stellar photons reaching them and this energy distribution is controlled by the absorption of the ions of the most common elements, hydrogen and helium.  

Close to the ionizing star hydrogen is ionized and oxygen is doubly ionized (easily traced by the \oiii\ 500.7 nm line). 
Helium is singly ionized (easily traced by the \heI\ 667.8 nm line). This is the \ozone\ zone. For the relatively cool Orion Nebula stars there is no higher ionization zone (where helium would be doubly ionized).

Further from the ionizing star and closer to the MIF is the narrow  \nzone\ zone, where helium is neutral and hydrogen is ionized. Nitrogen is singly ionized and is most visible in the  \nii\ 658.3 nm line. The easily visible Balmer \Ha\ and \Hb\ lines arise from both ionization zones.

\subsection{The basic approach}
\label{sec:basics}
We selected seven regions that were expected to illuminate the question of the ionizing star 
in various parts of the Orion Nebula. These all fall within the region of calibrated monochromatic images obtained with the MUSE \citep{weil15} multi-aperture spectrograph. Reddening corrected line ratios from the MUSE data-base are used throughout this study.
Figure~\ref{fig:mosaic} shows the location and sizes of these samples. 
They include regions near \tc\ and others on both sides of the Bright Bar.

The ratio of lines F(\oiii)/F(\Hb) and F(\heI )/F(\Ha ) within the \ozone\ zone and F(\nii)/F(\Ha) within the \nzone\ zone are dependent on the temperature of the ionizing star and a quantity defined as the ionization parameter \boldU~ \citep{agn3}, which is the ratio of the 
density of photons that can ionize hydrogen to the local hydrogen density (\NH). We can approximate the 
local hydrogen density as the electron density (\Ne) since hydrogen
 is essentially completely ionized and the $\approx 10\%$ contribution from \heI\ is small. 

 \begin{figure*}
\figurenum{2}
\plotone{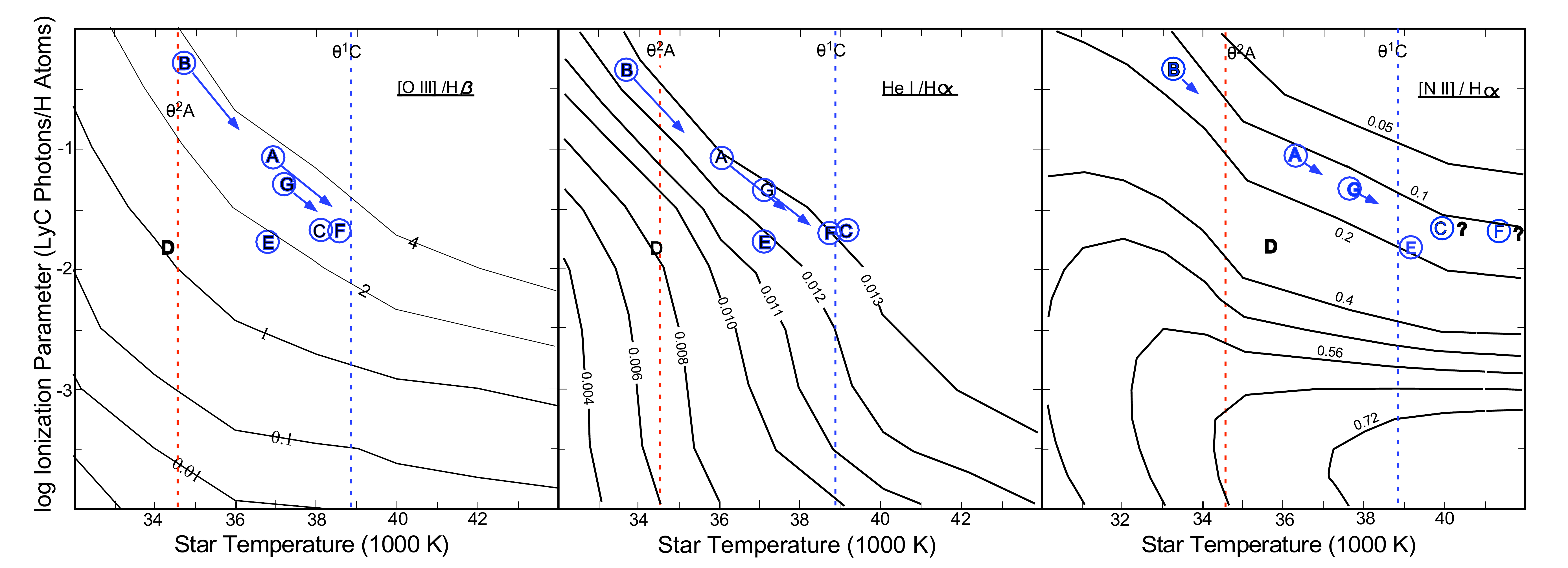}
\caption{
The ionization parameter versus ionizing star temperature is shown in all three panels, each depicting extinction insensitive line ratios. Dashed vertical lines indicate the temperatures of the two candidate ionizing stars. The segmented lines indicate the expected line ratio using Cloudy for varying values of the distance dependent ionization parameter assuming that the ionizing star is \tc. The seven samples from 
Figure~\ref{fig:mosaic} are shown. Those probably ionized by \tc\ are in blue and circled.
 Arrows indicate the direction of the correction to the ionization parameter for the
fact that it was calculated using the minimum possible distance from the star, i.e. the distance in the plane of the sky. The length of the arrow only
suggests the magnitude of that correction. Blue circles and lettering indicate samples most likely ionized by \tc.
}
\label{fig:ratios1C}
\end{figure*}

\begin{figure*}
\figurenum{3}
\plotone{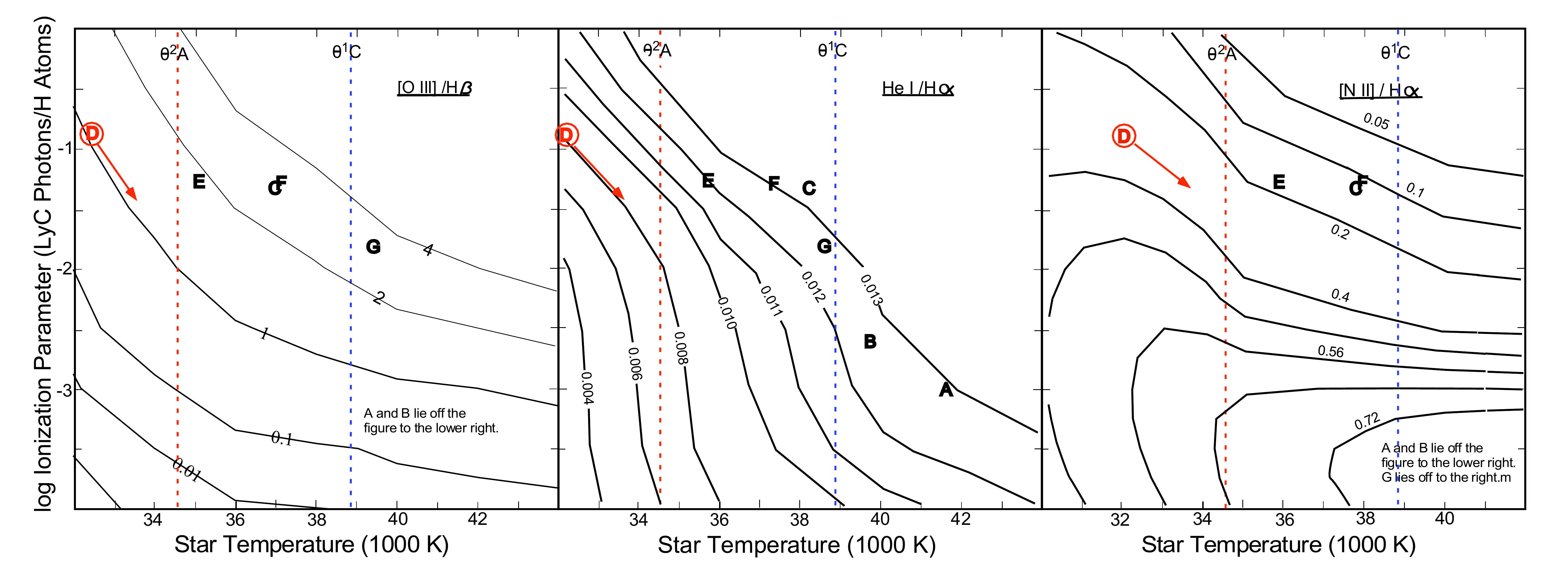}
\caption{
Like Figure~\ref{fig:ratios1C} except now the ionization parameters are calculated using distance from \tA. 
The samples probably ionized by \tA\ are in red and circled.
}
\label{fig:ratios2A}
\end{figure*}

 We can calculate the photon
 density (\rholyc) from the total ionizing photon luminosity of the ionizing star (\Qlyc) and the true distance to the emitting layer (\rtrue), and this leads to the relation

 \begin{equation}
 \boldU = 
 \frac {\rm Q_{LyC}}{\rm 4\pi~{\bf r}^2~c} ~\times~\frac{1}{\rm n\rm_{e}}
 \end{equation}
 
 where c is the velocity of light. 
 
 \subsection{Calculations}
 
 In this section we present a series of new ionization simulations of the Orion H~II region.
 We use version C17.00 of Cloudy, the spectral simulation code last described by \citet{gjf13}.
 The geometry is that used by \citet{BFM}.  
 It is assumed that the physical thickness of the H$^+$ layer that constitutes the H~II region is much thinner than the separation
 to the ionizing stars.  The geometry is plane parallel.
 The layer is assumed to be in hydrostatic equilibrium with the H$^+$ layer held back against the molecular cloud
 by the combination of gas pressure in the X-ray emitting hot gas surrounding the star cluster, and the absorbed outward momentum
 in the starlight.
 
 This model was originally applied to central regions of the Nebula, close to the Trapezium, where we are most likely viewing the
 H$^+$ layer nearly face on. However, the bright Huygens region is most likely bowl shaped, as described by \citet{gjf13} and \citet{wen95}.
 The observed surface brightness of the layer is affected by two things.
 First, as the distance of the star increases, the flux of ionizing photons, which sets the surface brightness, falls as \rtrue$^{-2}$.
 This causes the surface brightness to decrease with increasing radius.
 At the same time, the ``tilt'' of the H$+$ layer to our line of sight increases (see Fig 3 of \cite{wen95}) which causes the surface brightness
 to {\em increase} by a factor $cos(\theta)^{-1}$, where $\theta$ is the viewing angle measured relative to the normal to the layer,
 As long as our line of sight passes through the entire layer, the relative emission-line intensities {\em do not} change, only
 their total surface brightness.
 
 This simple geometry breaks down when the viewing angle approaches edge-on, since we then see successive regions projected
 on the plane of the sky, resolving the ionization structure.  
 This occurs within the Bar \citep{Sellgren} or near edges of proplyds. 
 In this case a more complex model can be developed
 \citep{Shaw}.
 For simplicity, in this study we use the \cite{BFM} approach and remain mindful that it will break down in edge-on cases.

 We  calculated the line ratios F(\oiii,500.7 nm)/F(\Hb), F(\heI,667.8 nm)/F(\Ha), and F(\nii,658.3 nm)/F(\Ha) for a variety of conditions. 
 The models adopted a progression of values of the ionizing star temperature and
 the ionization parameter. 
 We used the stellar SED described in \citet{Badnell}.
 In both figures the predicted values of the line ratios are plotted as contour lines. The lines in these ratios are at close wavelengths and relatively insensitive to reddening by foreground dust. Furthermore, the models were compared with reddening corrected line ratios.

\subsection{Methodology}
\label{sec:methodology}

By comparing the derived data for each sample with the expected values for the candidate ionizing star, we can identify that star. The apparent value of \boldU~ 
 (\Usky) was calculated from the 
 equation
 
 \begin{equation}
{\bf U}\rm_{sky} =
 \frac {\rm Q_{LyC}}{\rm 4\pi~r_{sky}^2~c} ~\times~\frac{1}{\rm n\rm_{e}}
 \end{equation}
 where \rsky\ is the separation of the sample and the candidate ionizing star and we have adopted the \Ne\ value derived from the red \sii\ ratio. This calculation was done for both \tc\ and \tA. This means that a given sample is represented twice, once in Figure~\ref{fig:ratios1C} and again in Figure~\ref{fig:ratios2A},
the first using \Usky\ for \tc\ and the second using \Usky\ for \tA, but the same line ratio for the sample.

 We do not know \rtrue, but we do know a lower limit, \rsky. Therefore an ionization parameter
 calculated using \rsky~(\Usky ) is an upper limit. If the assumed ionizing star is correct, then each derived point will lie  on the intersection of the observed line ratio and  \Tstar\ or to the upper left. A greater separation means a larger difference between \rtrue\ and \rsky, that is, there must be a greater distance correction.
 
If the observed point lies to the right of the intersection or if a projection along a constant line ratio does not reach the \Tstar~ value, then that sample is not ionized by the assumed star.  We do see two points (C and F) in the F(\nii)/F(\Ha) panel of Figure~\ref{fig:ratios1C} that lie to the right of the intersection with \Tstar\ for \tc. These fall in a very flat region for the predicted flux ratio. If we had adopted a smaller distance to the Orion Nebula, then the values of \Usky\ would be larger for all points, and Samples C and F would have shifted to near the intersection. However, our adopted value for the distance is already smaller than many modern determinations, as summarized in \citet{ode08a}. 

 It is more likely that the Cloudy model that we adopt is beginning to break-down in Sample C and Sample F
 due to the large viewing angle. Our model assumes
 that we are observing along a line-of-sight that goes through the entire H$^+$ layer. 
 When the MIF is highly tilted
 we resolve the ionization structure on the plane of the sky, and so sample along layers within the column.
 This compromises a comparison of a line-of-sight models against observations of a highly tilted region. We would expect that the results for the F(\nii)/F(\Ha) ratio would be affected the most since the [N II] line forms in a small part of the H$^+$ layer.  
 Although the peak of the tilt of the MIF is a maximum
 at the nearby Bright Bar, \citet{ode17} establish that the region near our Sample C and Sample F is already
 highly tilted.
 
 \subsection{Identifying the ionizing star from the observed ratios}
\label{sec:comparison}

In Figure~\ref{fig:ratios1C} we compare our observations with the brightest star in the Orion Nebula, \tc.
 Figure~\ref{fig:ratios2A} is similar, but assumes that the isolated bright star \tA\ is the source.
 
 In Figure~\ref{fig:ratios1C} we see that Samples A, B, C, F , and G are all ionized by \tc, although the F(\nii )/F(\Ha ) ratios for Samples C and F fall to the right of \Tstar (\tc). In F(\oiii)/F(\Hb) and F(\heI)/F(\Ha) we see that all samples are consistent with ionization by \tc, with the caveat of the remarks in Section~\ref{sec:methodology} about the high \Tstar\ values for Samples C and F. Taken at face value F(\oiii)/F(\Hb) and F(\heI)/F(\Ha) indicate that Samples C and F
 have spatial distances about equal to \rsky\ while Samples G , A, and B demand progressively larger corrections to the true distance. In the left and right panels it appears that Sample D could also be associated
 with \tc, but in the middle panel we see that no plausible correction for distance would make the line ratio compatible 
 with \Tstar\ for \tc.  All of the \tc\ associated samples fall within the central cavity of the nebula. Sample E can be associated with ionization by \tc\ with a small adjustment in U, which is allowable even though it is much further on the sky than either of the difficult to interpret Sample C or Sample F.
 
 In Figure~\ref{fig:ratios2A} we note that only Sample D agrees with ionization by \tA\ with similar distance corrections in all ratios. The appearance of the ionized edge of the nearby proplyd 244.3-439.8 indicates dominant ionization by \tA\ but with some possible contribution by \tc\ (Section~\ref{sec:proplyds}). Clearly Samples A, B, C, F, and G are not associated with \tA. Those samples lie on the \tc\ side of the Bright Bar and have already been associated with that star. Sample E agrees with an association with \tA~ in the F(\oiii)/(\Hb)
 without a distance correction; but, In the F(\heI)/F(\Ha) and F(\nii )/F(\Ha) ratios it lies too far to the right of \Tstar\ for \tA, which rules out \tA.
 
 The identification of the inner five samples as being ionized by \tc\ falls within the accepted model of this region of the Orion Nebula, i.e. that it is a concave irregular surface.The sample apparently ionized by \tA\ is consistent with 
 the Bright Bar and the regions southeast  from there being ionized by the nearest hot star. Sample E, although lying outside the
 Bright Bar is probably ionized by \tc, as discussed earlier in this section.
 \newpage
\begin{deluxetable*}{ccccccc}
\tablecaption{Ionization Parameters\tablenotemark{a} and Flux Ratios of the Seven Samples \label{tab:table1}}
\tablehead{
\colhead{Sample} &
\colhead{\Ne} \tablenotemark{b} &
\colhead{log~U(\tc)} &
\colhead{log~U(\tA)} &
\colhead{F(\oiii~500.7 nm)/F(\Hb)} &
\colhead{F(\nii~658.3 nm)/F(\Ha)} &
\colhead{F(\heI~667.8 nm)/F(\Ha)}
}
\startdata
A  &  6160 &-0.95 & -2.92 & 3.73 & 0.120 & 0.0129\\
B  & 4310  & -0.24 & -2.54 & 3.58 & 0.136    & 0.0126\\
C  & 2090  &-1.56 & -1.28 & 2.81 & 0.125    & 0.0132\\
D  & 1230  &-1.76 & -0.81 & 1.06 &  0.307    & 0.00812\\
E  &  1610 & -1.71 & -1.26 & 1.96 & 0.180    & 0.0118\\
F  & 1860 &-1.58 & -1.22  & 2.93 & 0.116    & 0.013\\
G  & 1970  &-1.27 & -1.70  & 3.03 & 0.129    & 0.0127\\
\enddata
\tablenotetext{a} {The ionization parameters are upper limits, being calculated from the distance to the ionizing star as projected on the sky.}
\tablenotetext{b} {Electron densities are in cm$^{-3}$ and are derived from the red [S II] doublet ratio using MUSE fluxes.}
\end{deluxetable*}

\section{Identifying the ionizing stars using observations of the proplyds}
\label{sec:proplyds}

There is a completely independent method of identifying the dominant source of ionization using high resolution images of the Orion Nebula proplyds. 
These are young stars with circum-stellar material  in or projected upon \hii\ regions. Under favorable orientations and locations the 
interior protoplanetary disks are rendered visible in silhouette. More relevant here, these disks are surrounded by much larger envelopes 
of gas and dust. A nearby hot star will form an ionization front in this gas  that is seen as a bright arc oriented towards the 
ionizing star. This allows one to identify the direction in the plane of the sky of the dominant ionizing star. 

In Figure~\ref{fig:mosaic} we have added white arrows on a representative sample of the proplyds. The arrows point
in the direction of the bright arc pointing towards that proplyd's ionizing star. The sampling is incomplete, but representative
of the proplyds closest to \tc, and includes all of the proplyds that indicate ionization by \tA.  
The latter four objects \citep{owh93,ow94}
are labeled with their names in the position-based designation system introduced by
 \citet{ow94} and later refined \citep{ode15}. The orientation of the bright arcs of those proplyds ionized by \tc\ are well illustrated in \citet{ode98}. Those four associated with \tA\ are shown in enlargements. It is to be noted that the circumstellar gas 
 of 196.6-426.6 is clearly ionized by both stars as may be the case for 244.3-439.8.

 \section{Using Spitzer Space Telescope Infrared Spectra Line Ratios}
\label{sec:rubin}
Published Spitzer Space Telescope observations of the infrared \neoniii\ 15.6 \micron\  and \neonii\ 12.8 \micron\ lines are also useful for determining the ionizing star. These are particularly useful because they both arise in the same \ozone\ zone and are unaffected by emission from the \nzone\ zone and any changes of abundances, as is the case for our optical line ratios. In the study of \citet{rubin} observations were made at eleven locations to the southeast of the central Orion Nebula. These samples all lie outside the Bright Bar, with the innermost being close to \tA. The progression of increasing distances is I4, I3, I2, I1, M1, M2, M3, M4, V1, V2, and V3. They all lie in a smooth region between the Bright Bar and the boundary of the Extended Orion Nebula (EON). Sample V2
falls near the edge of the sharp southeast boundary of the EON and Sample V3 falls on this boundary. These samples do not agree with the results from the others because Sample V3 is certainly an ionization front viewed edge-on and Sample V2 is probably a region that is rising towards that feature \citep{rubin} and thus our Cloudy models do not provide a guide to the conditions (Section~\ref{sec:methodology}). They are not similar to the other samples and are not included in our discussion of the general properties of the Spitzer samples. 

\begin{figure}
\figurenum{4}
\plotone{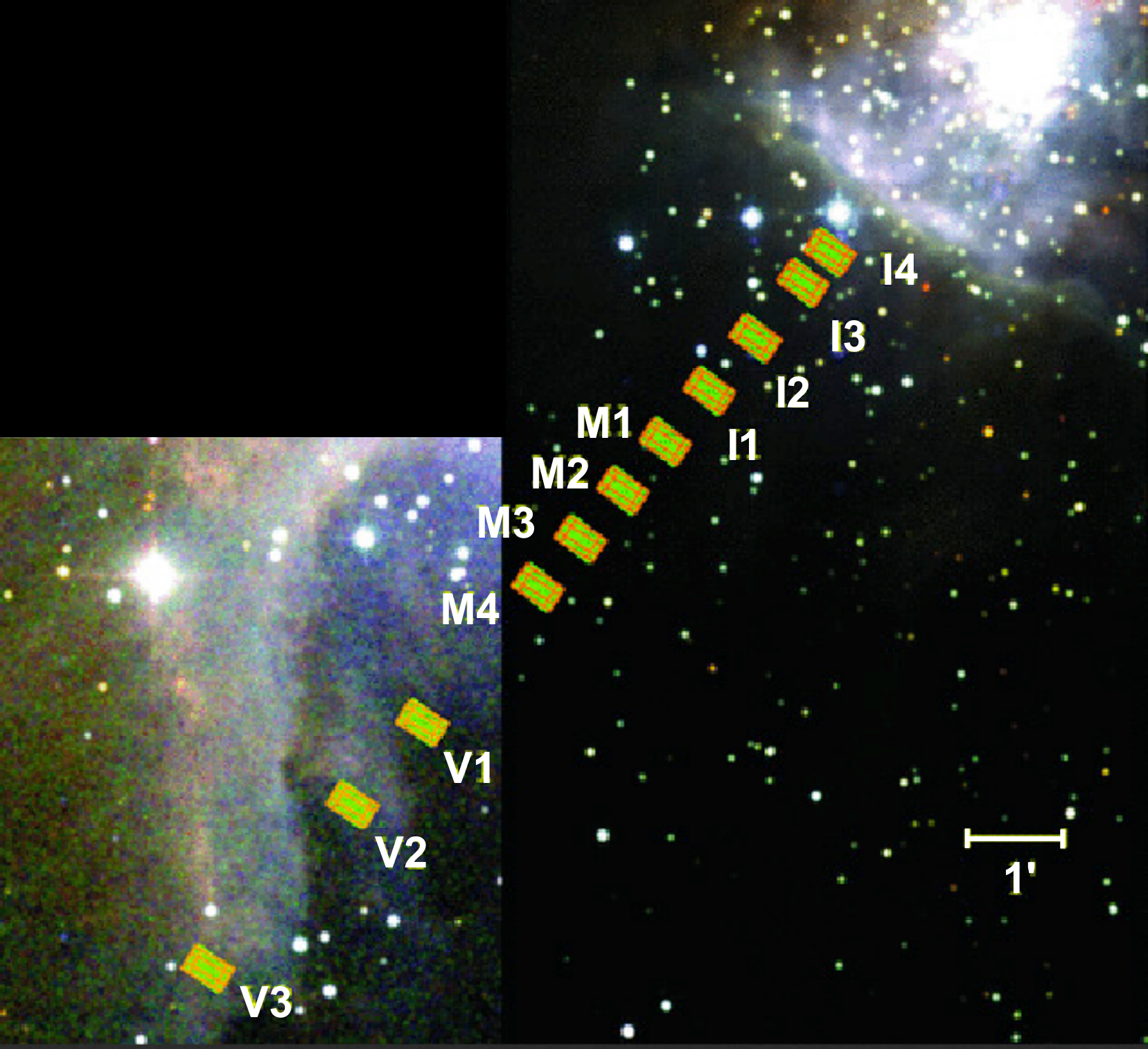}
\caption{
The Spitzer Samples are shown (after Figure 1 of \citet{rubin}) and are explained in the text. The east boundary of the EON appears at the lower left (southeast) and the Trapezium stars lie in the center of the bright northwest feature.}
\label{fig:rubin}
\end{figure}

\begin{figure}
\figurenum{5}
\plotone{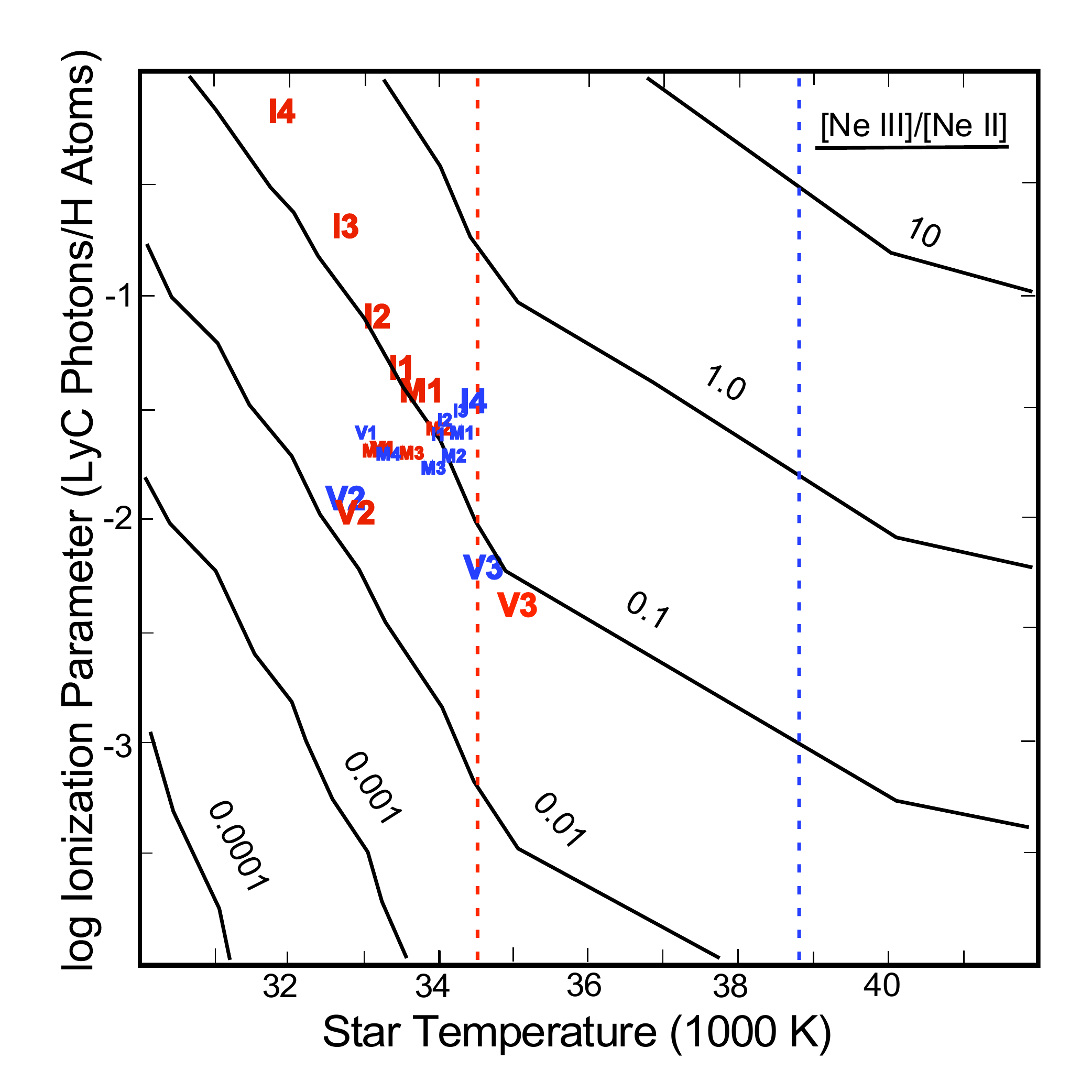}
\caption{
Like Figure~\ref{fig:ratios1C} and Figure~\ref{fig:ratios2A}  except now we use the ratio of the \neoniii\ 15.6 \micron\  and \neonii\ 12.8 \micron\ lines.
The symbols are samples from the study of \citet{rubin}  that begin close to \tA\ and proceed to the southeast with a maximum 
distance of 12.08 arcmin from \tc.  Colors indicate which star was assumed to be the ionizing source, blue for \tc\ and red for \tA. Samples V2 and V3 fall at and near the edge of the sharp southeast boundary of the Extended Orion Nebula and are not used in the discussion in Section\ \ref{sec:rubin}.}
\label{fig:neon}
\end{figure}

We have compared the neon line ratios with the predictions from the same set of Cloudy calculations used in the analysis of the optical line ratios.  We have used the \citet{rubin} \sii\ \Ne\ densities derived from overlapping optical observations for each sample.
The results are shown in Figure~\ref{fig:neon}.

Figure~\ref{fig:neon} shows that all of the samples that assume \tc\ to be the ionizing star (blue labels) cluster near \Tstar\ about 34000 K.
This is clear evidence that \tc\ does not ionize the region southwest of the Bright Bar.  To reconcile the observations with \Tstar\ for \tc\ 
would require all of the samples to have a large and similar correction for projection effects, even though their separations in the plane of the sky range from 2.6\arcmin\ to 12.08\arcmin. 

The comparison with the results assuming \tA\ (red labels) are fully consistent with that star as the dominant ionizing star, since Sample I4 is closest to \tA\ and our Sample D in 
the plane of the sky and the I3, I2, and I1 samples progress further away. Their locations in Figure~\ref{fig:neon} indicate a succession of 
		corrections for projection effects and at large distances the samples are close to \Tstar\ for \tA.
		
\section{Discussion}
\label{Discussion}

The results of Section~\ref{sec:ionization} and Section~\ref{sec:proplyds} show similar but not identical results. Some of the 
differences may be explained by the methodology.

The F(\oiii)/F(\Hb) and F(\heI)/F(\Ha) line ratios are indicators of the conditions in the \ozone\ zone as \oiii\ and \heI\ emission arise only there as does most of the \Ha\ emission. The gas density decreases \citep{ode01} with increasing
distance from the MIF, therefore most of the emission used in the F(\oiii)/F(\Hb) and F(\heI)/F(\Ha) line ratios method occurs near the onset of the 
\ozone\ zone, with a lesser \Ha\ contribution from the narrower \nzone\ zone. The distributions of doubly ionized oxygen (producing the \oiii\ emission) and \heI\ ions (producing the 667.8 nm emission) are slightly different within the \ozone\ zone. This means that 
the ratios using these two ions will give slightly different results.  In the  \nzone\ zone the \nii\ emission is well constrained, but the fact that the \Ha\ emission comes mostly from the \ozone\ zone, means that the F(\nii)/F(\Ha) ratio is intrinsically not as good a diagnostic.

In contrast, the proplyds give an indication of the ionization of objects located somewhere along the line of sight from
the foreground Veil and the MIF. The Orion Nebula Cluster is centered near the Trapezium and monotonically decreases
in stellar density with increasing distance in all directions. This means that a proplyd lying in the same direction as a 
ratios sample can lie well displaced towards the observer. This difference of sampling can explain the similarities and differences in the results.

\subsection{Location of the samples along the line of sight}
\label{sec:distances}

We can, however, approximately determine the position of the MIF samples with respect to the plane in the sky of the 
ionizing star. For each observed sample we can compare the ''observed" \Usky~with the value where the same line 
ratio intersects \Tstar. The shift in \boldU~gives us the square of \rtrue /\rsky, hence that ratio. For example, if the \Usky~is four times larger than where that line ratio intersects \Tstar, then \Usky~is four times too large and the true distance (\rtrue) is twice the separation in the plane of the sky 
(\rsky). It follows that the angle of \rtrue\ with respect to the plane of the sky ($\Theta$) will be
arccos(\rsky/\rtrue) and the distance along the line of sight (Z) is Z~=~\rsky$\times$tan\ $\Theta$.
What we actually determine is $\pm \Theta$ and $\pm$Z. 

Assignment to positive (towards the observer) or negative can usually 
be determined by comparing the Z values with respect to the quantitative models of the Orion Nebula. From the ratio of the
surface brightness in the radio continuum with that expected from ionization by \tc , both \citet{wen95} and \citet{hen05} 
derived a model of a concave surface with ridges (e.g. the one producing the Bright Bar). The former presumed a distance to the nebula
of 500 pc and the latter 430 pc.  Their derived separation (Z) values were -0.2 pc and about -0.15 pc. In their study of the spectrum of the substellar point using Cloudy, \citet{Badnell} used 0.10 pc. We assume in our discussion 
that the sub-\tc~distance is -0.15 pc. The line of sight positions of the samples can now be discussed in the order of increasing
distance from their ionizing stars.

\floattable
\begin{deluxetable}{cccccc}
\tablecaption{Positions of Samples with Respect to their Ionizing Stars \label{tab:table2}}
\tablehead{
\colhead{Sample} &
\colhead{Ionizing Star} &
\colhead{\rsky~(pc)} &
\colhead{\rtrue /\rsky} &
\colhead{$\Theta$} &
\colhead{Z~(pc)}
}
\startdata
A  &  \tc &0.054 & 1.77 & -60\arcdeg & -0.094 \\
B  & \tc  & 0.028 & 4.37 & -78\arcdeg& -0.134 \\
C  & \tc  &0.189 & 1.134   & 28\arcdeg    & 0.101    \\
D  & \tA  &0.060 & 2.81 & -72\arcdeg & - 0.180  \\
E  &  \tc & 0.254 &1.78 & 60\arcdeg & 0.444 \\
F & \tc   &0.179 & 1.134   & 28\arcdeg   & 0.096   \\
G & \tc  &0.139 & 1.53  & 55\arcdeg  & 0.196  \\
\enddata
\end{deluxetable}

\subsubsection{Samples ionized by \tc}
\label{sec:C1samples}

$\Theta$ for Sample A and B must be negative, otherwise samples this close to \tc\ would not be ionized by direct radiation.
Sample A would have Z~=~-0.09 pc and Sample B would have Z~=~-0.13 pc, the latter being comparable to the sub-\tc\ distance and
the former sample being 0.04 pc closer to the observer. This agrees with the rapid increase of surface brightness west of \tc.

If the values for Samples C, F, and G $\Theta$ were negative, then this would imply that the nebula is basically flat to the southeast from \tc. This is 
incompatible with the fact that the surface brightness in \Ha\ is nearly constant (which requires that the surface curves towards
the observer because the ionizing flux from \tc\ falls with \rtrue). These samples must have positive values of $\Theta$ and Z, which indicates
that the MIF surface has risen about 0.3 pc from the sub-\tc\ point over a distance of about 0.2 pc. 

Sample E certainly has positive values since it lies southeast of the Bright Bar, with this region being 0.44 pc towards the observer
from the plane of \tc\ and 0.6 pc above the sub-\tc\ point at a distance of 0.3 pc. Samples C, E, F, and G taken together indicate that the MIF rises steeply towards the observer to the southeast from \tc, which is the direction of the Bright Bar.

The region outside of the Bright Bar changes at about the position of the HH~203 and HH~204 shocks.
To the northeast of these shocks F(\nii)/F(\Ha) is high and F(\oiii)/F(\Hb) is low. This is reversed to the southwest
of these shocks and agrees with our Sample D and Sample E values. This transition is illustrated well in an F(\nii)/F(\Ha) 
image such as Figure~24 of \citet{ode15}. 

The southwest high ionization region is marked by numerous crenellations 
lying immediately southeast of the Bright Bar. These features have sharp boundaries and are likely to be shocks driven
by a series of outflows from the Orion-S star formation region \citep{ode15}.

If these shocks and earlier precursors are 
contributing to ionization in the region around Sample E, we can then accept that the sample does not give consistent results
for the distance correction for either for  \tc\ and it may be that \tA\ is playing a secondary role in ionizing this region.

\subsubsection{Samples ionized by \tA}
\label{sec:2Asamples}

The region between Sample D and \tA\ 
is relatively smooth and structure-free (except for the overlying shocks of HH~203 and HH~204).
This means that the sample must lie away from the observer relative to the plane of \tA. The progression of decreasing displacements of
the neon line ratios in going from Sample I4 to Sample I1 indicates that Sample I4 is well beyond \tA\ but by 
position I1 the ionization front is close to the same plane as \tA.

This result is quite different from that in \citet{rubin}. In that study they argue in a qualitative way
that all of the region along their samples (to the southeast, beyond the Bright Bar) is ionized by \tc, under the assumption that \tA\ is incapable of creating local \neoniii\ emission. Our Cloudy models indicate that this is 
not the case and our examination of the variations of the F(\neoniii)/F(\neonii) ratio firmly establish that \tA\ is the dominant ionizing star in this region.

In Section~\ref{sec:comparison} we established that Sample E, which lies outside the Bright Bar, is dominantly ionized by \tc. However,
this identification was good only for the F(\oiii)/F(\Hb) and F(\heI)/F(\Ha) ratios with a surprisingly large distance correction, while F(\nii)/F(\Ha) indicated that no distance correction was required. In the comparison
with line ratios predicated on ionization by \tA, the observed points were close to \Tstar\ for that star, but
always slightly hotter. It is most likely that Sample E is ionized by both \tc\ and \tA.

\section{Conclusions}
\label{sec:conclusions}

 We have confirmed the usual assumption that the bright portion of the Orion Nebula lying within
the boundary of the southeast Bright Bar is dominantly ionized by the hottest star in the Trapezium
grouping (\tc). With the exception of a single region (Sample E), all of the regions lying southeast of the 
Bright Bar are dominantly ionized by the much cooler isolated star \tA. Sample E is most likely ionized by both stars. This conclusion is verified by the orientation of the ionization boundaries of proplyds, with the exception of one proplyd (196.6-422.6)) that probably lies at a small spatial distance from \tA.

\acknowledgments

GJF acknowledges support by NSF (1108928, 1109061, and 1412155), NASA (10-ATP10-0053, 10-ADAP10-0073, and ATP13-0153), and STScI (HST-AR- 13245, GO-12560, HST-GO-12309, GO-13310.002-A, HST-AR-13914, HST-AR-14286.001 and HST-AR-14556). WK was supported in part by DFG grants KO 857/32-2 and KO 857/33-1.
 


\vspace{5mm}
\facilities{HST(WFPC2),  ESO(MUSE), SSL(IRS)}

\software{IRAF, Cloudy}

\end{document}